\pdfoutput=1

\documentclass[10pt,preprintnumbers,twocolumn,hyperref,nofootinbib,floatfix]{revtex4}
\usepackage{hyperref}
\usepackage{amssymb}
\usepackage{amsmath}
\usepackage{multirow}
\usepackage{mathrsfs}

\renewcommand{\bar}{\overline}
\renewcommand{\hat}{\widehat}

\begin{document}

\title{Effective Field Theories for\\ Local Models in F-Theory and M-Theory}
\author{Jacob L. Bourjaily}
\affiliation{Department of Physics, Princeton University, Princeton, NJ 08544}
\affiliation{School of Natural Sciences, Institute for Advanced Study, Princeton, NJ 08540}%??

\begin{abstract}
Requiring a strictly local origin of visible sector phenomenology is perhaps the strongest, most falsifiable condition that one can impose on string theory at the high scale: it at once excludes a vast majority of the string landscape, and yet leads naturally to constructions that can be surprisingly realistic (and familiar). Yet only for local models can gravity be made parametrically weak while keeping the strength of gauge- and Yukawa-couplings fixed---a limit which is well-motivated by low-energy experiments. Conveniently, the entire class of high-scale effective field theories that can arise from such local models in F-theory and M-theory can be classified according to simple, purely group-theoretic rules. In this note, we describe these rules from the viewpoint of an effective field theorist with little interest in the underlying geometry or high-scale physics, and we discuss the general predictions these models have for low-energy phenomenology.\vspace{-1.4cm}
\end{abstract}

\maketitle

%%%%%%%%%%%%%%%%%%%%%%%%%%%%%%%%%%%%%%%%%%%DRAFT COMMAND%%%%%%%%%%%%%%%%%%%%%%%%%%%%%%%%%%%%%%%%%%%
%\draft
%%%%%%%%%%%%%%%%%%%%%%%%%%%%%%%%%%%%%%%%%%%DRAFT COMMAND%%%%%%%%%%%%%%%%%%%%%%%%%%%%%%%%%%%%%%%%%%%
\vspace{-0.45cm}

\section{Introduction}\vspace{-0.5cm}

The existence of a strict effective-field-theoretic limit in which the interactions involving Standard Model fields can be parametrically decoupled from the effects of quantum gravity places surprisingly strong---even predictive---constraints on any UV-completion of the Standard Model in the context of \mbox{F-theory} or \mbox{M-theory} \cite{Beasley:2008dc,Beasley:2008kw,Bourjaily:2009vf}. These constraints turn out to exclude an enormous fraction of the string landscape, leaving only a tiny patch which, surprisingly, can \mbox{(only just)} accommodate some of the most realistic string models to date \cite{Bourjaily:2009vf,Heckman:2008jy,Heckman:2008ads,Marsano:2008jq,Heckman:2008qa,Blumenhagen:2008aw,Donagi:2008ca,Donagi:2008kj,Wijnholt:2008db,Heckman:2009bi,Donagi:2009ra,Marsano:2009ym,Bouchard:2009bu} including those with gauge coupling unification, doublet-triplet splitting, gauge-mediated supersymmetry breaking, matter-parity, an axion solution to the strong CP-problem, and even semi-realistic Yukawa textures. Although these models have been exhaustively described in terms of high-scale geometry \cite{Bourjaily:2009vf}, it should be emphasized that the rules by which they are classified are essentially group-theoretic and do not themselves require any familiarity with \mbox{\mbox{F-theory}} or \mbox{\mbox{M-theory}} to use. The purpose of this paper is to present the basic rules for building these models as effective field theories in a way that bypasses the underlying high-scale physics, requiring instead only a modest familiarity with representation theory---such as that of unified model building.

The reason why decoupling turns out to be so constraining---and hence predictive---is that in the limit of $M_{Pl}\to\infty$, any physics involving sectors separated by the bulk will decouple; therefore, the requirement that the visible-sector can remain intact as an effective theory in this limit requires that virtually all of the physics relevant to phenomenology can arise {\it locally} in the internal geometry. In \mbox{F-theory} and \mbox{M-theory}, the local structures that give rise to non-abelian gauge symmetries and interacting massless charged matter are well understood and easily classified in terms of ALE-fibrations \cite{Acharya:2001gy,Acharya:2004qe,Berglund:2002hw,Bourjaily:2008ji,Bourjaily:2007kv,Katz:1996xe,Witten:2001uq}. And because of the close correspondence between the geometry of ALE-spaces and the structure of simply-laced (`ADE') groups---$SU_n(\equiv A_{n-1})$, $SO_{2n}(\equiv D_n)$, and $E_n(\equiv E_n)$---this classification directly translates into simple, group-theoretic rules for determining which effective field theories can result from these purely local structures in \mbox{F-theory} or in \mbox{M-theory}.

There are two principal reasons for choosing to forego the underlying high-scale physics and describe these models as effective field theories built using just group-theoretic rules. First, the number of phenomenologists familiar with the esoterica of group theory (especially as applied to unified-model building), is much greater than the number of those familiar with the esoterica of \mbox{F-theory} or \mbox{M-theory}; and so merely translating what is known in \mbox{F-theory} and \mbox{M-theory} into the `more common tongue' of group theory can greatly broaden the search for better models. Secondly, for those already fluent in \mbox{F-theory} and \mbox{M-theory}, a precise statement of the rules as they are known can be viewed as a challenge for further refinement: there are still many important questions that remain unsettled in the literature, and precise claims naturally encourage illustrative objections. When we state the precise rules in \mbox{Section \ref{model_rules}}, we will make sure to distinguish between those which are truly established, and those which are to some extent still conjectural or may be subject to locally invisible, global consistency requirements such as tadpole cancellation.%\vspace{-0.025cm}

We can roughly state the rules for `locally engineerable' models as follows. For any simple ADE-group $G$ and any of its subgroups \mbox{$H\times U_1^1\times\cdots\times U_1^k\subset G$}, there exist purely local models in \mbox{F-theory} and \mbox{\mbox{M-theory}} which have gauge-symmetry group $H$ and a spectrum of massless chiral matter coming entirely from the branching of the adjoint of \mbox{$G\to H\times U_1^1\times\cdots\times U_1^k$}; by this we mean that each vector-like pair of representations \mbox{$\mathbf{R}\oplus\bar{\mathbf{R}}$} in the adjoint branching will contribute either $\mathbf{R}$ {\it or} $\bar{\mathbf{R}}$ to the massless spectrum. Furthermore, the geometry of such a model will generate {\it every} \mbox{$(H\times U_1^1\times\cdots\times U_1^k)$}-invariant operator involving these fields in the \mbox{superpotential}.%\vspace{-0.025cm}
\newpage
Perhaps importantly, it turns out that models in \mbox{\mbox{F-theory}} are considerably less rigid than what we have described so far; indeed, \mbox{F-theory} will actually allow for any multiple of $\mathbf{R}$ or $\bar{\mathbf{R}}$ from the branching $G\to H$ to appear in the massless spectrum, including the possibility that neither appears. This gives \mbox{F-theory} an enormous flexibility---a flexibility which is not shared by \mbox{M-theory}. In contrast, unless we impose specific (geometric) constraints, \mbox{M-theory} {\it must} include {\it either} $\mathbf{R}$ {\it or} $\bar{\mathbf{R}}$ in the massless spectrum. But having one copy of each representation from the branching $G\to H$ is often incompatible with the needs of phenomenology---such as the absence of local non-abelian anomalies. Therefore, the constraints by which some matter representations can be excluded in \mbox{\mbox{M-theory}} play a very important role; and ultimately, these constraints are responsible for making models in \mbox{M-theory} enormously more constrained than their \mbox{F-theory} cousins.%\looseness=-1
\vspace{-0.4cm}

\vspace{-0.05cm}
\section{Locally Engineerable Models}\label{model_rules}\vspace{-0.35cm}
\subsection{Locally Engineerable Geometries for \mbox{F-Theory} and \mbox{M-Theory}}\vspace{-0.35cm}
Both in \mbox{F-theory} and in \mbox{M-theory}, gauge-fields are understood to arise locally from co-dimension four orbifold-type singularities, and massless charged matter to arise from places along these singularities where the type of orbifold singularity is enhanced \cite{Katz:1996xe}. In \mbox{F-theory}, co-dimension four singularities lie along complex two-cycles, and can be enhanced along embedded complex curves \mbox{(`matter-curves')} which, because fields along them live in \mbox{5+1} dimensions, can generate chiral matter only if parity is broken, e.g. by internal magnetic flux \cite{Beasley:2008dc,Beasley:2008kw}. In \mbox{M-theory}, co-dimension four singularities are real three-cycles, and these can be enhanced at isolated points; because these points live in \mbox{3+1} dimensions, they manifestly support chiral matter \cite{Acharya:2001gy,Acharya:2004qe,Cvetic:2001nr,Witten:2001uq}. It is useful to note that geometries with these structures can be described in a way that does not depend on whether we are discussing \mbox{F-theory} or \mbox{M-theory} \cite{Bourjaily:2009vf}.

In Ref.\ \cite{Bourjaily:2009vf} we described how any Calabi-Yau four-fold or $G_2$-manifold constructed as an ALE-fibration will naturally contain the singular structures necessary to generate gauge theory with charged, massless chiral matter. Furthermore, we showed that these manifolds automatically possess the larger structures which connect disparate matter singularities necessary to generate interactions in the superpotential. In \mbox{F-theory}, cubic couplings are generated by the triple-overlap of chiral wave functions living along mutually intersecting matter-curves; and in \mbox{M-theory}, they are generated by Euclidean M2-brane instantons wrapping supersymmetric three-cycles which support multiple conical matter singularities. Although these structures are intuitively topologically-non-generic, we showed in \mbox{Ref.\ \cite{Bourjaily:2009vf}} that they are ubiquitous features of any ALE-fibred compactification manifold.
\newpage
We can summarize the geometric results of \mbox{Ref.\ \cite{Bourjaily:2009vf}} as follows. Let $G$ be any simple\footnote{If $G$ were not simple, then each of its simple components would decouple in the limit of \mbox{$M_{Pl}\to\infty$}.} ADE-group, and consider any of its maximal subgroups \mbox{$H\times U_1^1\times\cdots\times U_1^k\subset G$} with \mbox{$k\geq2$};\footnote{For \mbox{M-theory}, it is only strictly necessary that $k\geq1$; but this does not appear to lead to any useful novelties for \mbox{M-theory}.} let us choose to write the branching of the adjoint of $G$ into this subgroup in the form\footnote{The branching of ADE adjoints is described in generality, with several pedagogical examples in Section 2 of Ref.\ \cite{Bourjaily:2009vf}. For a more thorough discussion, we recommend \mbox{Refs.\ \cite{FultonRepTheory,Slansky:1981yr}}.} \begin{equation}
\mathrm{adj}\left(G\right)=\mathrm{adj}\left(H\times \prod_{j=1}^kU_1^j\right)\bigoplus_{\vec{q}\in Q^+}\left(\mathbf{R}_{\vec{q}}\oplus\bar{\mathbf{R}}_{\text{-}\vec{q}}\right),\label{gen_branching}
\end{equation} where $Q^+$ is the space of $k$-vectors $\vec{q}$ of $U_1$-charges whose first non-vanishing entry is positive. (Notice that this fixes a convention associating the `un-barred' component representation of each vector-like pair $\mathbf{R}\oplus\bar{\mathbf{R}}$ to be the one with `positive' $U_1$-charges.) It is a useful fact of representation theory that each representation $\mathbf{R}_{\vec{q}}$ in (\ref{gen_branching}) is {\it uniquely} identified by its charge-vector $\vec{q}$. Another useful fact is that invariance of an operator under \mbox{$(U_1^1\times\cdots\times U_1^k)$} implies invariance under $H$ as well.\footnote{Both of these statements follow from the fact that the charges $\vec{q}$ label sets of roots in the weight lattice of $\mathrm{adj}(G)$; see \mbox{Section 3} of Ref.\ \cite{Bourjaily:2009vf} for details.} Letting $\hat{G}$ denote the ALE-space of type $G$, {\it there exists an explicit, local, $\hat{G}$-fibred Calabi-Yau four-fold (for use in \mbox{F-theory}) and a similarly constructed $\hat{G}$-fibred $G_2$-manifold (for use in \mbox{M-theory}), which have} 
\begin{enumerate}
\item a co-dimension four singularity of type $H$, generating a gauge-symmetry group $H$;
\item  a matter singularity of type $\mathbf{R}_{\vec{q}}$ for every charge $\vec{q}\in Q^+$, except perhaps those of any chosen subset $K\subset Q^+$ and those in the space spanned by $K$;\footnote{Notice that if $\mathrm{dim}(\mathrm{span}(K))=\mathrm{dim}(Q^+)=\,$the number of $U_1$-factors$\,=k$, then no matter singularities will be present in the manifold.}\label{item2}
\item and the structures necessary to generate every \mbox{$(H\times U_1^1\times\cdots\times U_1^k)$}-invariant operator among these fields in the superpotential. 
\end{enumerate}
Using the tools of Ref.\ \cite{Bourjaily:2009vf}, it is not hard to actually construct these manifolds. And doing so will allow one to identify all the local moduli of the geometry which determine all of the operators in the effective superpotential.\footnote{This is only strictly true in the case of \mbox{M-theory}, where the complete form of the coefficients' dependence on the local moduli is well understood \cite{Bourjaily:2009vf}; in \mbox{F-theory}, the wave functions living along matter-curves need not be normalized within the local patch under consideration, and so it is not generally possible to actually compute coefficients of operators in the superpotential, despite knowing all the local moduli.}

%\newpage
Notice that property (\ref{item2}.) describes which matter singularities can be chosen to be absent from the local geometry via what is known as {\it parallel projection}. This is possible because it turns out that the {\it location} of each matter-singularity is fixed by its $U_1$-charges---namely, each field $\mathbf{R}_{\vec{q}}$ is supported at the solution to an equation of the form $\vec{q}\cdot\vec{f}(W)=0$, where $W$ is the internal manifold.\footnote{Notice that that this implies that any pair of representations in the branching (\ref{gen_branching}) with linearly independent $U_1$-charges will be {\it geographically separated}.} By adjusting some of the maps $\vec{f}$ to be non-vanishing constant maps, some of the fields can be forced out of the local geometry. Such geometric constraints will play an important role in \mbox{M-theory}, because every matter singularity in the geometry is expected to give rise to massless matter---which is often undesirable. 

Although less rigorously understood, it seems likely that there also exist local manifolds based on any quotient of the space of $U_1$-charges of the form $Q^+/\Gamma$.\footnote{It is sometimes the case that the equivalence $\Gamma$ will enhance the rank of $H$---e.g.\ some will result in a symmetry group \mbox{$(H'\times U_1^1\times\cdots\times U_1^{k-1})$} such that \mbox{$H'\supset H\times U_1^k$}. Imposing such a quotient leads to nothing novel (being the same as having started with a higher-rank symmetry group $H'$).} Specifically, for any equivalence relation $\Gamma$ of $Q^+$, it is not hard to construct a local geometry with a matter singularity for every $\mathbf{R}_{\vec{q}}$ with $\vec{q}\in Q^+/\Gamma$ (except perhaps those spanned by any given subset $K\subset Q^+/\Gamma$), and the structures to generate every $(U_1\times\cdots\times U_1^k)/\Gamma$-invariant operator among these fields.\footnote{Because $(\prod_j U_1^j)$-invariance implies $H$-invariance, we strongly suspect that this continues to be the case for $(\prod_j U_1^j)/\Gamma$.\vspace{-0.3cm}} This has the geometric interpretation of identifying various two-cycles in the ALE-fibres. Several examples illustrating this type of construction were presented in \mbox{Ref.\ \cite{Bourjaily:2009vf}}. One point regarding these quotients that remains to be adequately understood is that they almost invariably lead to higher-rank matter-singularities in the geometry. Although there exist well-understood examples of multiply-enhanced singularities in \mbox{M-theory} \cite{Acharya:2001gy}, they less understood in general. 

%\vspace{-0.115cm}
Although the local geometries that we can engineer for F-theory have essentially the same structures as those for M-theory, the actual field theories that result in each case can be quite different. The most important source of the difference between the two frameworks is the flexibility of choosing the massless matter spectrum in F-theory. %\mbox{M-theory} lacks this flexibility, making it much more rigid about what it predicts. 
But we should emphasize that even if the same spectrum of massless matter were realized in \mbox{F-theory} as in \mbox{M-theory}, the two UV-completions would be quite different: in \mbox{F-theory}, Yukawa couplings are fixed by wave-function overlap integrals, while in M-theory they are set by the action of Euclidean M2-brane instantons, which are typically very much more hierarchical. These differences can have important phenomenological consequences.\nopagebreak

%\newpage
%\vspace{-0.6cm}
\subsection{Massless Matter in F-Theory From Fluxes}%\vspace{-0.4cm}
Because matter singularities in \mbox{F-theory} live in 5+1 dimensions, they naturally generate matter in `$\mathcal{N}=2$ hypermultiplets' (and are ergo massive) upon compactification to 3+1 dimensions. In order to obtain chiral matter, one must add internal magnetic flux along some of the matter-curves.\footnote{More specifically, fluxes are added along each of the complex two-cycles whose intersection defines the matter-curve; these are just (possibly exceptional) branes in the type IIb description.} These fluxes are naturally quantized, and are almost always separately tunable\footnote{It is possible that not all the matter-curves are distinct. This will be the case whenever there are two or more proportional charge-vectors in $Q^+$. For example, for \mbox{$E_7\to SU_5\times SU_2\times U_1^a\times U_1^b$}, $\mathbf{133}\supset (\mathbf{10},\mathbf{2})_{1,0}\oplus(\bar{\mathbf{5}},\mathbf{1})_{2,0}$. In this case, both representations would live along the same curve, and it is not clear if it is possible to choose $n_{1,0}$---the number of massless $(\mathbf{10},\mathbf{2})_{1,0}$'s---independently of $n_{2,0}$---the number of massless $(\bar{\mathbf{5}},\mathbf{1})_{2,0}$'s. In \mbox{M-theory}, such independent choices always appear to be possible---subject to the condition that $n_{\vec{q}}=\pm1$ .}, leading to any multiple\footnote{Negative numbers of $\mathbf{R}$'s should be understood as net $\bar{\mathbf{R}}$'s.}, say $n_{\vec{q}}$, of the representation $\mathbf{R}_{\vec{q}}$ in the massless spectrum. Importantly, adding no flux along a given matter-curve will result in no net chiral matter in \mbox{3+1} dimensions, making the need for parallel projections superfluous: exotic, unwanted matter singularities are easily ignored in the effective field theory. 

Therefore, the geometric results above imply the following algorithm for constructing models in \mbox{F-theory}:

\begin{enumerate}
\item choose any branching \mbox{$G\to H\times U_1^1\times\cdots\times U_1^k$} and write down a table of resulting representations $\mathbf{R}_{\vec{q}}$ together with their corresponding \mbox{$U_1$-charges} $\vec{q}$;\\
\mbox{~}\hspace{0.15cm}\parbox{7.3cm}{(if desired, take a quotient of the space of charges $Q^+$, rewriting the charges $\vec{q}$ accordingly)}
\item for each $\mathbf{R}_{\vec{q}}$ in the table, choose any integer $n_{\vec{q}}\,$ (possibly zero) to be in the massless spectrum;
\item write all $H\times(U_1^1\times\cdots\times U_1^k)$-invariant operators among the massless fields in the superpotential. 
\end{enumerate}

A substantially less-explicit freedom which is frequently employed in the literature to construct phenomenological models in \mbox{F-theory} is to imagine that some matter-curves which locally appear distinct are in fact the same globally. In effective field theory, this would correspond to the ability to identify any pair of matter representations $\mathbf{R}_{\vec{q}_a}$ and $\mathbf{R}_{\vec{q}_b}$ that are equivalent under $H$ (but not \mbox{$U_1^1\times\cdots\times U_1^k$}). However, while this seems plausible from an {\it ultra}-local viewpoint, we expect there to be rigid topological obstructions to such identifications in general. For the present, we recommend that such glue be used cautiously, as such models may be globally inconsistent.

\subsection{Manifestly Massless Matter in M-Theory}\vspace{-0.4cm}
In stark contrast to \mbox{F-theory}, matter singularities in \mbox{M-theory} are co-dimension seven, and therefore support fields that automatically live in 3+1 dimensions, manifestly generating $\mathcal{N}=1$ chiral multiplets \cite{Acharya:2001gy,Atiyah:2001qf,Witten:2001uq}. Because there are no tunable fluxes involved in obtaining chiral matter, there is no (known) analogue in \mbox{M-theory} of \mbox{F-theory}'s ability to generate multiple copies of a given representation at a single singularity, or of having some matter singularities in the geometry which are capable of {\it not} generating chiral matter. 

In general, each matter singularity of type $\mathbf{R}_{\vec{q}}$ present in the geometry will generate $n_{\vec{q}}=\pm1$ copies of $\mathbf{R}_{\vec{q}}$ in the massless spectrum. Therefore, unwanted matter must be either parallel-projected out of the geometry through some choice of $K\subset Q^+$, or must somehow be dynamically projected out of the spectrum at high energies. Of the two, parallel projection is both easier and more general. However, it is worth noting that each separate condition used to project-out matter (counted by $\mathrm{dim}(\mathrm{span}(K))$) reduces the moduli space of the resulting geometry, causing most models in \mbox{M-theory} to end up having very few adjustable parameters. 

As with \mbox{F-theory}, these basic facts of the underlying theory and the geometric considerations mentioned above lead directly to a simple procedure for building local models in \mbox{M-theory}:

\begin{enumerate}
\item choose any branching \mbox{$G\to H\times U_1^1\times\cdots\times U_1^k$} and write down a table of resulting representations $\mathbf{R}_{\vec{q}}$ together with their corresponding \mbox{$U_1$-charges} $\vec{q}$;\\
\mbox{~}\hspace{0.15cm}\parbox{7.3cm}{(if desired, take a quotient of the space of charges $Q^+$, rewriting the charges $\vec{q}$ accordingly)}
\item choose any set of unwanted representations $\left\{\mathbf{R}_{\vec{k}}\right\}$ for $\vec{k}\in K$ and strike these from the table together with every field $\mathbf{R}_{\vec{q}}$ whose charge \mbox{$\vec{q}\in\mathrm{span}(K)$};
\item for each representation that remains in the table, choose to retain either it or its complete conjugate in the massless spectrum (that is, choose $n_{\vec{q}}=\pm1$);
\item write all $H\times(U_1^1\times\cdots\times U_1^k)$-invariant operators among the massless fields in the superpotential.
\end{enumerate}

An important caveat for \mbox{M-theory} is that it remains to be demonstrated in the literature that {\it all} conjugation choices $n_{\vec{q}}=\pm1$ can be made independently of one another. In \mbox{F-theory}, this choice is manifest: the $U_1$-fluxes supported along each matter curve are separately tunable because they involve distinct branes. In \mbox{M-theory}, the issue is much more subtle, and the claim we make here is even in direct conflict with an earlier claim in the literature \cite{Bourjaily:2008ji}. Nonetheless, at least for examples simple-enough to have direct type IIa duals in terms of intersecting branes, it is not hard to verify that each $n_{\vec{q}}$ can indeed be chosen independently of the others. This issue will be explored in greater detail in \mbox{Ref.\ \cite{Bourjaily:2009aa}}.

\section{General Features of Local Models in F-Theory and M-Theory}\vspace{-0.4cm}
\subsection{Three Generations and $E_8$-Structure}\vspace{-0.3cm}
Because the clay out of which local models are molded in \mbox{F-theory} and \mbox{M-theory} is the branching of an adjoint $G\to H$, it is clear that the size and structure of the starting group $G$ will strictly limit the variety of matter representations and the ways in which they can interact in the resulting model. Perhaps surprisingly, it turns out that essentially no group smaller than $E_8$ can generate the complete variety of couplings necessary to describe the Standard Model\footnote{Conveniently, there is reason to expect that only $\hat{E_8}$-fibrations are compactifiable: there is only one compact Calabi-Yau two-fold, namely $K3$, and it has inside of it two copies of $\hat{E_8}$ (see e.g.\ Ref.\ \cite{Aspinwall:1995vk}).}---and this is true even for a single generation \cite{Bourjaily:2009vf}. And because there is no larger group that contains it, $E_8$ is the essentially unique starting point for any phenomenological model. 

Because of this, familiarity with the branching of the adjoint of $E_8$ is especially useful for local model building. Unfortunately however, most of the familiar reference tables on the branching of group representations \mbox{(e.g., \cite{Slansky:1981yr})}  often ignore the $U_1$-charges which are critical for us here. And so, for the convenience of the reader we have chosen to include the complete branching of \mbox{$E_8\to E_6\to SO_{10}\to SU_5\to SU_3\times SU_2\times U_1^Y$} in \mbox{Table \ref{E8_branching}} at the end of this note. 

As emphasized in Refs.\ \cite{Bourjaily:2009vf,Bourjaily:2007vx}, a very suggestive feature of $E_8$ is that its adjoint naturally branches to {\it three generations} of matter. This can be understood from the branching \mbox{$E_8\to E_6\times U_1^a\times U_1^b$}:\vspace{-0.135cm}
\begin{align}\vspace{-1cm}
\hspace{-0.25cm}\mathbf{248}=\!\!\!&\phantom{\oplus}(\mathbf{78}_{0,0}\oplus\mathbf{1}_{0,0}\oplus\mathbf{1}_{0,0})\label{e8_to_e6}\\&\oplus\mathbf{27}_{\phantom{\text{-}}1,0}\oplus\mathbf{27}_{0,\phantom{\text{-}}1}\oplus\bar{\mathbf{27}}_{\phantom{\text{-}}1,\phantom{\text{-}}1}\oplus\mathbf{1}_{\phantom{\text{-}}2,\phantom{\text{-}}1}\oplus\mathbf{1}_{\phantom{\text{-}}1,\phantom{\text{-}}2}\oplus\mathbf{1}_{\phantom{\text{-}}1,\text{-}1}\nonumber\\
&\oplus\bar{\mathbf{27}}_{\text{-}1,0}\oplus\bar{\mathbf{27}}_{0,\text{-}1}\oplus\mathbf{27}_{\text{-}1,\text{-}1}\oplus\mathbf{1}_{\text{-}2,\text{-}1}\oplus\mathbf{1}_{\text{-}1,\text{-}2}\oplus\mathbf{1}_{\text{-}1,\phantom{\text{-}}1}.\nonumber\vspace{-0.25cm}
\end{align}
$\!\!\!$(Notice that according to the conventions set by (\ref{gen_branching}), the `un-barred' representation $\mathbf{R}_{1,1}$ is actually $\bar{\mathbf{27}}_{1,1}$ because it has positive $U_1$-charges---unlike \mbox{$\mathbf{27}_{\text{-}1,\text{-}1}\left(\equiv\bar{\mathbf{R}}_{\text{-}1,\text{-}1}\right)$}.)

As we have described, there exists local $E_6$-models in \mbox{F-theory} with massless matter for any choices $n_{\vec{q}}$ for each of the vector-like pairs \mbox{$\mathbf{R}_{\vec{q}}\oplus\bar{\mathbf{R}}_{\text{-}\vec{q}}$} in (\ref{e8_to_e6}), and models in both \mbox{M-theory} and \mbox{F-theory} if the $n_{\vec{q}}$ are restricted to be $\pm1$ \mbox{(or possibly $0$)}. One set of such choices, for example, will lead to a model whose renormalizable\footnote{Notice that at the non-renormalizable level, operators such as $\mathbf{27}_{1,0}\mathbf{27}_{0,1}\mathbf{27}_{0,1}\mathbf{1}_{\text{-}1,\text{-}2}$ are also allowed. These can be entirely understood as arising from integrating out massive modes connected by cubic operators. For example, $\mathbf{27}_{1,0}\mathbf{27}_{0,1}\mathbf{27}_{0,1}\mathbf{1}_{\text{-}1,\text{-}2}$ can be generated by $(\mathbf{27}_{1,0}\mathbf{27}_{0,1}\left(\bar{\mathbf{27}}_{1,1}\right)^{\dag})\left(\bar{\mathbf{27}}_{1,1}\mathbf{27}_{0,1}\mathbf{1}_{\text{-}1,\text{-}2}\right)$ upon integrating out a massive KK-mode $\left(\bar{\mathbf{27}}_{1,1}\right)^{\dag}\bar{\mathbf{27}}_{1,1}$.} superpotential is of the form, \begin{equation}W=\lambda_1\,\mathbf{27}_{1,0}\mathbf{27}_{0,1}\mathbf{27}_{\text{-}1,\text{-}1}+\lambda_2\,\mathbf{1}_{2,1}\mathbf{1}_{\text{-}1,\text{-}2}\mathbf{1}_{\text{-}1,1}.\label{w_e6}\end{equation}
Recall that each $\mathbf{27}$ of $E_6$ represents an entire generation of matter---including its own Higgs fields (and their coloured partners), a right-handed neutrino, and one other Standard Model singlet.\footnote{For a review of unification and the role of $E_6$, see e.g.\ \cite{Langacker:1980js,Hewett:1988xc}.} And if $E_6$ were somehow Higgsed down to $SO_{10}$, $SU_5$, or even all the way to $SU_3\times SU_2\times U_1^Y$, the single cubic coupling of $\mathbf{27}$'s in (\ref{w_e6}) would branch to all the familiar interactions of an $E_6$-like GUT, still retaining the accidental $\mathbb{Z}_3$-flavour symmetry of (\ref{w_e6}). Of course, without any additional structure, these interactions would include those that lead to rapid proton decay, flavour-changing neutral currents, and other phenomenological disasters. Therefore, we are naturally motivated to consider constructing local models with less gauge symmetry at the high-scale, such as ones based on $SO_{10}$, $SU_5$, or even \mbox{$SU_3\times SU_2\times U_1^Y$} gauge symmetry. 
\vspace{-0.6cm}

\subsection{Geometric Unfolding vs. Grand Unification} \vspace{-0.4cm}
Of course, one is naturally led to ask: what is the difference between starting with a local model based on $E_8\to SO_{10}$ and simply breaking a local $E_6$-model to one with $SO_{10}$ symmetry via some Higgs mechanism? After all, the possible matter content of a model based on \mbox{$E_8\to SO_{10}$} could be obtained by simply branching each representation of (\ref{e8_to_e6}) according to \mbox{$E_6\to SO_{10}\times U_1^c$}:\footnote{It should be noted that $U_1^c$ is normalized to be nothing but $U_1^{PQ}$, the familiar Peccei-Quinn symmetry.}
\vspace{-0.1cm}
\begin{align}\vspace{-0.5cm}\mathbf{78}=\,&(\mathbf{45}\oplus\mathbf{1})\oplus\bar{\mathbf{16}}_{3}\oplus\mathbf{16}_{\text{-}3};\nonumber\\
\hspace{-1.25cm}\mathrm{and}\quad\mathbf{27}=\,&\mathbf{16}_{1}\oplus\mathbf{10}_{\text{-}2}\oplus\mathbf{1}_{4}.\label{e6_to_so10}\vspace{-0.5cm}\end{align}

\vspace{-0.2cm}

\noindent Therefore, a model based on \mbox{$E_8\to SO_{10}$} would appear to share all the structure that we would have expected from a model with $E_6$ broken by adjoint Higgses.\footnote{Of course, the branching of the $\mathbf{78}$ contributes an additional $\bar{\mathbf{16}}_{0,0,3}$ to the spectrum. However, this additional matter can easily be parallel-projected out of the spectrum by taking \mbox{$\vec{q}=(0,0,3)\in K$.} }

Nevertheless, it is important to emphasize that a model based on \mbox{$E_8\to SO_{10}$} does not {\it a priori} have any dynamical connection to one based on \mbox{$E_8\to E_6$}. Although the `unfolding' of $E_6$ to $SO_{10}$ would appear smooth locally, such a transition would almost certainly be forbidden in a compact geometry: `unfolding' is the geometric version of smoothly separating stacks of D-branes, which is always possible in a non-compact, local construction, but almost never possible in a compact model. And although it is useful to know the structure of Higgs-induced-branching (because representation theory is context-independent), there is no need for a dynamical restoration of symmetry at high energies: for example, if we had started with a model based on \mbox{$E_8\to SU_3\times SU_2\times U_1^Y$}, the high-scale theory would generate only the Standard Model's \mbox{$SU_3\times SU_2\times U_1^Y$} gauge symmetry, without grand-unification being manifest or mandatory.

But perhaps the most important way in which models which are `geometrically unfolded' out of high-rank singularities differ from models with sequential degrees of unification is the freedom to choose either $\mathbf{R}$ or $\bar{\mathbf{R}}$ to be present in the low-energy spectrum---the ability to `conjugate' any of the possible matter representations in the model. Such a freedom is extremely unusual from the viewpoint of traditional unified model building and can only be understood in terms of a higher-dimensional theory\footnote{The essential origin of this flexibility is that the local geometries we construct in \mbox{F-theory} and \mbox{M-theory} are based on ALE-fibrations, and ALE-spaces are naturally `$\mathcal{N}=2$-like' (because they are Calabi-Yau two-folds, and hence hyper-K\"ahler). This is especially clear in \mbox{F-theory}, where matter singularities generate hypermultiplets in 5+1-dimensions---and so only generate chiral fields in 3+1 if there is non-vanishing internal magnetic flux.}. And yet, this freedom of \mbox{F-theory} and \mbox{M-theory} can play an extremely important role phenomenologically, and has several important applications.% in model building. 

A simple example of how such conjugations can be used to achieve phenomenological goals can be seen already in the example $E_8\to SO_{10}$. Ordinarily, the branching of a single $\mathbf{27}$---as in (\ref{e6_to_so10})---would generate terms in the superpotential including those of the form\vspace{-0.3cm}\begin{equation}W\supset \mathbf{10}_{\text{-}2}\mathbf{10}_{\text{-}2}\mathbf{1}_{4};\vspace{-0.3cm}\end{equation}
this term could give rise to a familiar \mbox{`dynamical-$\mu$'}-like operator $S_\mu H^u H^d$. But it is sometimes preferable\footnote{This is the case, for example, in the extremely realistic models described by \mbox{Refs.\ \cite{Heckman:2008ads,Marsano:2008jq,Ibe:2007km}}, realized concretely in \mbox{Ref.\ \cite{Bourjaily:2009vf}}.} to exclude this operator entirely from the superpotential, and choose instead to dynamically generate $\mu$ through a Giudice-Massiero-like operator of the form $\left(\mathbf{1}_{\text{-}4}\right)^\dag\mathbf{10}_{\text{-}2}\mathbf{10}_{\text{-}2}$ in the K\"ahler potential \cite{Giudice:1988yz} . Notice that either choice of the conjugation of $\mathbf{1}_{4}$ will allow one of these operators while simultaneously excluding the other. 

Another example of how the ability to conjugate fields in the low-energy spectrum can naturally permit solutions to many of the problems of traditional grand-unified models is as follows. If we were to construct a high-scale model with gauge symmetry group \mbox{$SU_3\times SU_2\times U_1^Y$}, then we could use this freedom to exclude all of the lepto-quark and di-quark operators\footnote{Recall that these operators can be seen to descend from the $\mathbf{27}^3$ interaction of $E_6$. Letting $(D,H^u)\subset\mathbf{5}_{H^u}$ and $(D^c,H^d)\subset\bar{\mathbf{5}}_{H^d}$ denote the Higgs multiplets with their coloured partners, the lepto-quark operators are of the form $Q\,D^c\,L$, $D\,d^c\,\nu^c$, and $D\, u^c\, e^c$, and the di-quark operators are of the form $Q\,Q\,D$ and $D^c\,u^c\,d^c$ \cite{Langacker:1980js}. Conjugating $D$ and $D^c$ will cause all such operators to be non-gauge invariant.} that would otherwise lead to rapid proton decay by simply conjugating each of the Higgs fields' coloured partners. Of course, this is {\it not} doublet-triplet splitting, since by starting with mere \mbox{$SU_3\times SU_2\times U_1^Y$} gauge symmetry at the high scale, the doublet- and triplet-components of the Higgs $\mathbf{5}$ and $\bar{\mathbf{5}}$'s are manifestly distinguished. But the ability to keep the (conjugates of the) Higgs' coloured partners in the low-energy spectrum without inducing proton decay reflects the powerful new flexibility offered by local models in \mbox{F-theory} and \mbox{\mbox{M-theory}}.
 
But doublet-triplet splitting---that is, ensuring that the triplet partners of some Higgs fields are absent from the low-energy spectrum---is not only useful to ensure proton stability, but it is also critical for gauge-coupling unification: it is well known that the Standard Model gauge couplings can unify at high energies (ironically) only if the spectrum includes {\it incomplete GUT multiplets}. Recall that gauge coupling unification in the MSSM requires that the both of the Higgs doublets appear appear without their $SU_5$ partners in the low-energy spectrum. Therefore, in order for a grand-unified model to be consistent with known low-energy data, it must have a mechanism to generate {\it incomplete} representations at the high scale. 

Although beyond the scope of our present discussion, it turns out that natural mechanisms exist in both \mbox{F-theory} and \mbox{M-theory} to achieve doublet-triplet splitting at the high scale. We will not have more to say about how this works in this paper, but it is worth mentioning here that it is almost always possible to arrange a unified gauge group to be broken in such a way that incomplete multiplets survive at low energies. 
\vspace{-0.2cm}
\subsection{Sparsity of the Superpotential}\vspace{-0.2cm}
One of the most important and general features of any model constructed according to the rules described above is the overall {\it sparsity} of the superpotential that results. That these models have so few operators in the superpotential would appear to be highly non-generic from a bottom-up, effective field theory point of view: it is natural to expect every operator to appear in the action unless explicitly forbidden by symmetries; and there is a natural prejudice against adding any non-compulsory symmetries by hand. But in the context of local models, such additional symmetries {\it are} compulsory: {\it every local model will include approximate global, discrete symmetries descending from a group of higher-rank than the gauge-symmetry}; and these additional symmetries {\it do not require a cascading tower of sequential unifications} to explain their existence.

Importantly, although local models exist for the branching of any branching \mbox{$G\to H$}, the only group $G$ with enough structure to generate every type of interaction needed for phenomenology is $E_8$. Therefore, not only do local models predict the existence of new, highly-restrictive symmetries, but they fix precisely {\it which} additional symmetries are present: they must be those which descend from $E_8$. For any model in the `standard' chain of unification \mbox{$E_8\to E_6\to SO_{10}\to SU_5\to SU_3\times SU_2\times U_1^Y$}, these additional $U_1$ symmetries are those given in \mbox{Table \ref{E8_branching}}. In that Table, $U_1^a$ and $U_1^b$ are the diagonal generators of a ``broken'' $SU_3$ flavour-symmetry; $U_1^c$ is nothing but the familiar Peccei-Quinn symmetry from $E_6\to SO_{10}\times U_1^c$; and $U_1^d$ is the symmetry coming from the branching $SO_{10}\to SU_5\times U_1^d$. Notice that because matter parity is a subgroup of $U_1^c$, all local models will have this as an approximate, global discrete symmetry.

Although a bottom-up effective field theorist may find all these extra $U_1$-symmetries quite constraining \mbox{(they are)}, it is worth noting that from the top-down it is rather surprising that any operators are generated in the superpotential at all! Indeed, the structures needed to generate interactions are topologically non-generic (this is especially so in \mbox{M-theory}\footnote{Superpotential interactions in \mbox{M-theory} are generated by Euclidean M2-brane instantons living along supersymmetric three-cycles which support multiple conical singularities. For a field at a given singularity to appear multiple times in the superpotential---e.g.\ $H^u$---it must lie along multiple, mutually intersecting supersymmetric three-cycles; but three-cycles don't generically intersect in a seven manifold at all! And so, the proof in Ref.\ \cite{Bourjaily:2009vf} that {\it any} multiply-unfolded, ALE-fibred $G_2$-manifold will have these structures, is quite remarkable.}), and so it is quite remarkable that there exist local models which have great enough a variety of couplings to be even modestly phenomenological, let alone truly realistic.

While these additional symmetries are helpful for model building, they also lead to some generic difficulties. For example, the sparsity of Higgs couplings na\"ively complicates the possibility of obtaining any realistic spectrum of quark masses. But as pointed out in \mbox{Ref.\ \cite{Heckman:2008qa}}, this problem can be naturally remedied through higher-dimensional, effective operators\footnote{This is similar to the familiar Froggatt-Nielen mechanism \cite{Froggatt:1978nt}.} involving Standard Model singlets which acquire large vacuum expectation values;\footnote{There are also other mechanisms to flesh-out the Yukawa matrices, such as self-intersecting matter-curves in \mbox{F-theory}; but because even the most generic higher-dimensional operators work well enough in most circumstances, it is difficult to justify any thing more extreme.} this point was also made in the context of explicit examples in \mbox{Ref.\ \cite{Bourjaily:2009vf}}. Indeed, such singlet vevs are completely natural from the UV-perspective\footnote{Non-Abelian-singlet vevs are naturally generated by Green-Schwarz-induced D-terms in the scalar potential, which are generated for each locally-anomalous $U_1$-symmetry.}, and generally lead to a more densely populated---although quite hierarchical---superpotential. It is worth noting that such hierarchies are extremely well-motivated by low-energy data. These kinds of problems are also seen to arise in the neutrino sector, and are likely to have similar solutions (see e.g.\ Ref.\ \cite{Bouchard:2009bu}).

Another important, general feature of these models results from the fact that each representation $\mathbf{R}_{\vec{q}}$ in the massless spectrum is {\it uniquely} identified\footnote{In \mbox{F-theory}, multiple `generations' of $\mathbf{R}_{\vec{q}}$ can appear in the massless spectrum; this will not affect our conclusions about the superpotential.} by its $U_1$-charge vector $\vec{q}$, and that if $\mathbf{R}_{\vec{q}}$ is in the massless spectrum, then $\bar{\mathbf{R}}_{\text{-}\vec{q}}$ is not. This forbids any fundamental quadratic operator from appearing the superpotential\footnote{The only left-handed, chiral field with opposite $U_1$ charges to $\mathbf{R}_{\vec{q}}$ is $\bar{\mathbf{R}}_{\text{-}\vec{q}}$, and both cannot be massless simultaneously. This claim can be violated globally in models not capable of being written locally as a single ALE-fibration, such as in \mbox{Ref.\ \cite{Bilal:2003bf}}. But when this is the case, the coefficient of any interaction between the fields would be on the order of $e^{-2\pi/\alpha_{GUT}}$; and this is the limit of precision of any claim based solely on local data.\label{bilalref}}, automatically solving the `$\mathcal{O}(0)$ $\mu$-problem' of supersymmetry: a {\it bare} $H^uH^d$ operator cannot be generated in any local model. The same argument also forbids the existence of any pure Majorana masses for neutrinos. At best, such quadratic operators can be generated effectively through higher-dimension operators involving fields which acquire vacuum expectation values. However, it is important to note that precisely which effectively quadratic operators can arise in this way is strictly determined by $U_1$-invariance. 

In \mbox{M-theory}, because at most there can be exclusively a single copy of $\mathbf{R}_{\vec{q}}$ {\it or} $\bar{\mathbf{R}}_{\text{-}\vec{q}}$ in the massless spectrum\footnote{This may seem somewhat tautological: if $\mathbf{R}_{\vec{q}}$ and $\bar{\mathbf{R}}_{\text{-}\vec{q}}$ were both in the massless spectrum, then a high-scale mass term would be allowed and the fields would be projected out of the spectrum. This is not our argument. Rather, because in local models the {\it geography} of each matter singularity is dictated by its $U_1$-charges, and these uniquely determine each representation in the branching $G\to H\times U_1^1\times\cdots\times U_1^k$, there simply do not exist local models with geographically-isolated fields having the same or opposite $U_1$-charges. This need not be the case for compact models, as was mentioned in footnote (\ref{bilalref}).}, the only field with the opposite quantum numbers to $\mathbf{R}_{\vec{q}}$ is its CP-conjugate $(\mathbf{R}_{\vec{q}})^{\dag}$. This means that the only gauge-invariant quadratic terms in the K\"ahler potential are canonical. This fact could potentially be important for gravity-mediated supersymmetry breaking scenarios such as those of Refs.\ \cite{Acharya:2006ia,Acharya:2008zi,Acharya:2007rc}, because a supergravity spurion---a singlet under the visible-sector---would only able to generate {\it flavour-diagonal} soft-masses. Although flavour-universality would also be needed to cure all problems related to flavour-changing neutral currents, this is a good improvement on what is ordinarily thought to be problematic in gravity mediation scenarios (see e.g. Ref.\ \cite{Chung:2003fi} for a review). 

\vspace{-0.4cm}
\section{Parametrical Input from the UV}\vspace{-0.3cm}
As we have seen, much of the phenomenology of local models can be entirely understood in terms of the possible spectra of massless, charged matter: because matter-fields are determined (up to conjugation) by the branching of ADE-adjoint representations, all local models will feature additional $U_1$-symmetries which have general consequences for low-energy phenomenology; and because the overall conjugation of each field in the massless spectrum can be freely adjusted, many interesting phenomenological scenarios are possible. Combining the familiar $U_1$-structure which descends from $E_8$ with the ability to conjugate fields in the spectrum, leads to effective theories that are quite reminiscent of but decidedly distinct from those encountered in traditional unified model building. 

What we have described so far has been the most essential data of an effective field theory: the spectrum of massless matter, and how each field transforms under the symmetries of the theory. But in order to study these models in more detail, it is useful to have at least a qualitative understanding of the parametrical data which descends from the underlying, high-scale physics. In keeping with our general philosophy, therefore, in this Section we will continue to avoid any detailed discussion of the UV-physics. We suggest that readers interested in a more thorough discussion should consult the more technical literature of F-theory \mbox{(e.g. \mbox{Refs.\ \cite{Beasley:2008dc,Beasley:2008kw}})} and \mbox{M-theory} (e.g. \mbox{Ref.\ \cite{Acharya:2004qe}}).

\subsection{Hierarchies and Scales}\vspace{-0.3cm}
One of the most appealing aspects of locally-engineered models is that they naturally lead to very large hierarchies within the superpotential. This is partially due to the sparsity of the superpotential, as we have described. But much more importantly, it turns out that the mechanisms responsible for generating interactions in F-theory and M-theory both will generically lead to relatively large hierarchies among the non-vanishing operators in the superpotential. It is worth stressing that large hierarchies are in fact observed in Nature; and so, this (albeit {\it a posteriori}) prediction of locality should be considered quite encouraging. 

In both \mbox{M-theory} and \mbox{F-theory}, the coefficients of operators in the superpotential can range over several orders of magnitude; and yet for similar models, the hierarchies generated in M-theory should be much more pronounced than those observed in F-theory. This merely reflects the differences in how interactions are generated in the two frameworks: in M-theory, interactions arise via Euclidean M2-brane instantons, and so operators are generically exponentially suppressed by an instanton action;\footnote{Parametrically, the instanton action is roughly the volume of the supersymmetric three-cycle which the M2-brane wraps, as measured in string units.} while in F-theory, interactions are generated by the mutliple-overlap of wave functions for fields supported along mutually-intersecting matter-curves. Therefore, superpotential coefficients should be distributed over an exponential range in M-theory, while in F-theory they are expected to vary as a result of wave-function fall-off and should typically be on the scale of a geometric suppression---reflecting the fact that the matter wave functions are be spread-out over entire matter-curves.

Although it would be difficult today to claim that either M-theory or F-theory {\it predicted} a large Yukawa coupling for the top-quark, either framework can {\it naturally accommodate} such a coupling. Furthermore, if an $\mathcal{O}(1)$ Yukawa coupling for one generation were to be imposed by hand, then either framework would predict there to be very large hierarchies separating this generation from the remaining two For example, consider a model in F-theory for which each generation lives along a separate set of matter-curves. Then in order for the top-quark's Yukawa coupling to be $\mathcal{O}(1)$, the up-type Higgs field's wave function must be highly-peaked at the triple-intersection with the top-quark. But in order for the Higgs field's wave function to be peaked at its intersection with the top-quark, it must fall off very rapidly away from that intersection, forcing the other up-type quarks to have hierarchically small Yukawa couplings. The analogous argument for M-theory was presented in some detail in \mbox{Ref.\ \cite{Bourjaily:2009vf}}.

Another appealing feature of local models is the very small number of fundamental scales that are needed to define the theory. It is {\it almost} true that $M_{GUT}$ is the only energy scale in the model, but $M_{GUT}$ can only be meaningfully fixed at a finite value if also given $M_{Pl}$: after all, if the limit \mbox{$M_{Pl}\to\infty$} were taken rigorously, then any asymptotically-free effective field theory would be best described in the far UV, with \mbox{$\alpha_{GUT}\to0$}. So more specifically, local models must be endowed with both an energy scale $M_{GUT}$, and a small parameter, \begin{equation}\epsilon\equiv\frac{M_{GUT}}{M_{Pl}}.\end{equation} Numerically, the fact that $\epsilon\sim 10^{-3}$ is very convenient for perturbation theory, and leads to phenomenologically encouraging numerology.\footnote{Although beyond the scope of our present discussion, it is possible to argue the plausibility of obtaining many of the phenomenologically-important scales near $M_{GUT}$ by combining appropriate powers of $\epsilon$ with reasonable hierarchies among operators in the superpotential---including, in particular, the scales of supersymmetry-breaking and its mediation, right-handed neutrino masses, and the breaking of Peccei-Quinn symmetry \mbox{(see e.g.\ Ref.\ \cite{Heckman:2008ads})}.} It is worth mentioning that although local model building requires $\epsilon\ll1$ for consistency, this hierarchy may itself be dynamically quite natural for string compactifications---especially in the \mbox{type IIb}-limit of F-theory.\footnote{See, for example, \mbox{Refs.\ \cite{Conlon:2008cj,Conlon:2008wa}.}}

%Geometrically, $M_{Pl}$, $M_{GUT}$, and $\alpha_{GUT}$ are related by \begin{equation}\frac{1}{\alpha_{GUT}(M_{GUT})}\sim \frac{\mathrm{vol}(\hat{H})}{\ell_{*}^{d-8}}\sim\frac{\ell_{GUT}^{d-8}}{\ell_*^{d-8}},\label{voleqn}\end{equation} where $\mathrm{vol}(\hat{H})$ represents the ($(d-8)$-dimensional) internal volume of the co-dimension four singularity $\hat{H}$ and $\ell_*$ is (essentially) the string scale, which can directly related to the four-dimensional Planck length---with details depending on whether one is discussing F-theory or M-theory.\footnote{It is difficult to give a more precise form of (\ref{voleqn}) because in a complete theory we expect (\ref{voleqn}) to be corrected significantly by e.g. threshold effects near the string scale (see e.g.\ Ref.\ \cite{Conlon:2009xf}).}

\newpage
\subsection{Gauged and Global Symmetries: Anomalies, and Axions}
We have seen that local models manifestly include several additional $U_1$-symmetries; it is natural to wonder why these symmetries can be treated any differently from the non-Abelian symmetry group $H$. The only principle difference between the $U_1$-symmetries and $H$ is that---for purely phenomenological reasons---we will always require $H$ to be gauged at low energies. Notice that the rules we described in Section \ref{model_rules} would seem to allow for arbitrarily-bad anomalies to be present in the effective theory at high energies. 

Of course, the presence of anomalies in an effective field theory need not be any cause for concern: they merely indicate (or necessitate) the existence of a UV completion for the theory at a scale where new degrees of freedom are present to cancel the anomalies \cite{Preskill:1990fr}. And of course, local models in F-theory or M-theory come naturally-equipped with such a cutoff: the string scale or Planck scale.\footnote{The string scale and Planck scales can (and typically do) differ parametrically in local models, and it requires a more detailed analysis to determine which is the relevant scale to use as the cutoff in a given context.} And the UV-degrees of freedom necessary to ultimately cancel these anomalies are axions participating in the Green-Schwarz mechanism. Therefore, the presence of an anomaly for any of the symmetries \mbox{$H\times U_1^1\times\cdots\times U_1^k$} merely indicates that the symmetry is broken at relatively high energies, surviving at low energies as an approximate, global discrete symmetry. 

Before we go any further, however, it is necessary to mention one possible caveat to any local analysis of anomalies in F-theory or M-theory. For any local model {\it other than those descending from $E_8$}, the massless spectrum of matter obtained using the rules above is not manifestly {\it complete}: it is possible that there exists additional matter charged under \mbox{$H\times U_1^1\times\cdots\times U_1^k$} which are geographically-separated\footnote{In M-theory, geographic isolation can mean virtually-complete decoupling (except via gauge-interactions) because any Yukawa coupling involving the original, local fields, and the new, geographically-separated fields will be suppressed action of instantons wrapping cycles roughly at least as large as the original local region. For any phenomenologically-complete model in M-theory, this means that interactions with non-local matter would be suppressed at least on the order of about $e^{-2\pi/\alpha_{GUT}}$. } from the spectrum of matter singularities obtained from the branching \mbox{$(E_8\neq) G\to H\times U_1^1\times\cdots\times U_1^k$}. The reason why models descending from $E_8$ do not share this possibility is that the existence of another matter singularity outside of the local patch would require the existence of a simple \mbox{rank-9} ADE-group containing $E_8$.\footnote{See Ref.\ \cite{Bourjaily:2009vf} for a more detailed discussion.} At any rate, the presence of anomalies in any massless spectrum obtained according to the rules outlined in Section \ref{model_rules} either requires the existence of additional, {\it geographically-separated} massless fields in the spectrum to cancel the anomaly, or will cause the symmetry to be broken to a discrete, global subgroup at high energies.\footnote{The possible existence of additional, massless charged matter also implies that having an anomaly-free local spectrum of matter is by itself neither necessary nor sufficient to ensure that a symmetry be gauged at low-energies. As before, statements about completeness can be made much stronger for any model based on an $\hat{E_8}$-fibration.}

If the local spectrum of matter is taken as complete, then it is straight-forward to calculate the scale of symmetry breaking for any anomalous $U_1$-symmetry \cite{Preskill:1990fr}. It is not hard to see\footnote{Equations (\ref{cubic_anom}) and (\ref{mixed_anom}) are easy to derive as corrections to the boson propagator through a diagram obtained by gluing-together two triangles.} that a pure-gauge (cubic) anomaly would generate a vector mass-squared for the $U_1$-vector field parametrically of the order\begin{equation}
\mu^2\sim g^2\frac{1}{16\pi^2}\left(\frac{g^2\mathrm{Tr}(Q^3)}{16\pi^2}\Lambda\right)^2;\label{cubic_anom}\end{equation} while a mixed gravitational-gauge anomaly would generate a mass-squared of the order\begin{equation}\mu^2\sim g^2\frac{1}{16\pi^2}\left(\frac{\mathrm{Tr}(Q)}{16\pi^2}\frac{\Lambda^3}{M_{Pl}^2}\right)^2,\label{mixed_anom}\end{equation} where $\Lambda$ is the scale at which the Green-Schwarz mechanism turns on to cancel the anomaly in the UV. From equations (\ref{cubic_anom}) and (\ref{mixed_anom}) it is easy to see that the symmetry-breaking scale induced by an anomaly in the spectrum will always be parametrically below the cutoff scale of theory---typically separated by three or four orders of magnitude.\footnote{This is true simply from the powers of $16\pi^2$ in the denominators of (\ref{cubic_anom}) and (\ref{mixed_anom}) . Even greater hierarchies are possible if the gauge-coupling were made sufficiently small.} Because $\epsilon=\tfrac{M_{GUT}}{M_{Pl}}\sim 10^{-3}$, this can easily allow for an anomalous $U_1$-symmetry to be broken at a scale substantially below $M_{GUT}$. 

Such anomalies could be quite important phenomenologically; for example, it could be used to engineer a Peccei-Quinn solution to the strong CP-problem. Recall that the Peccei-Quinn symmetry naturally descends from $E_6$, and was listed in \mbox{Table \ref{E8_branching}} as $U_1^c$. If the massless fields of a local model descending from $E_8$ were conjugated in such a way as to ensure a non-vanishing anomaly for $U_1^c$, then the Peccei-Quinn symmetry would be broken at a scale which could be parametrically below $M_{GUT}$, and possibly low-enough to be consistent with cosmological bounds without fine tuning. This possibility was argued to be realized for the models described in \mbox{Refs.\ \cite{Heckman:2008ads,Heckman:2008jy}}. 

Importantly, because $U_1$-invariance dictates which operators are generated locally in the superpotential, any instantons which violate these $U_1$'s must involve the global geometry, and therefore lead to $U_1$-violating effects suppressed on the order of $e^{-2\pi/\alpha_{GUT}}$. These effects are exceedingly small, and imply that each $U_1$ will survive as at least a discrete, global approximate symmetry to very low energies. In particular, it is not implausible for the leading instanton-contribution to the the PQ axion potential to be from QCD-instantons, allowing for a true axion solution to the strong CP-problem. 

The ease with which we can solve the strong CP-problem with a Peccei-Quinn axion, however, also implies a possible generic problem with these models. If we had chosen a spectrum for which there were several anomalous $U_1$-symmetries, there may be too many light axions available at the QCD-scale. Only one linear-compbination of these fields can play the role of the QCD axion, while the others will remain as light degrees of freedom well-below the QCD scale. Such additional, light scalars could lead to undesirable cosmological effects.
%
%\subsection{Supersymmetry Breaking and its Mediation}
%The breaking of a $U_1$-symmetry through the Green-Schwarz mechanism can also be understood as generating a field-dependent, effective `FayetÐIliopoulos'-like \mbox{D-term} to the scalar potential\footnote{Being dynamical, field-dependent terms, they should arguably not be called Fayet-Iliopoulos operators \cite{Komargodski:2009pc}.}, which can generate supersymmetry breaking. 

\section{Conclusions and Outlook}\vspace{-0.3cm}
Even without knowing the precise origins of moduli stabilization or fully understanding which local models can be realized in compact geometries, we find that locality itself places severe restrictions on the class of effective theories that can be realized in \mbox{F-theory} and \mbox{M-theory}. Nature may or may not admit such a strict decoupling limit (let alone have a UV-completion understood at all in terms of string theory), but this condition represents perhaps the most concrete and falsifiable hypothesis that one can make in string theory, naturally encouraging a thorough exploration of its consequences.% \newpage

%Local models can indeed be excluded quite easily at the LHC or other TeV-probing experiments in the near future. For example, it is virtually impossible to construct models with strongly-coupled  hidden- or dark-sectors that are coupled by messengers to the Standard Model: because \mbox{$SU_5\times SU_3\times SU_2\times U_1\subset E_8$} is a maximal subgroup---$SU_3$ always playing the role of a flavour-symmetry---$SU_2$ is essentially the largest possible hidden-sector available for either supersymmetry breaking or explaining dark matter. 

%{\it more conclusions\ldots}
%\ldots 

%\newpage
%\vspace{-0.5cm}
\acknowledgements\vspace{-0.3cm}
This work was originally inspired by conversations with Nima Arkani-Hamed, Gary Shiu, and Paul Langacker. We are especially grateful to Nima Arkani-Hamed, Paul Langacker, Joe Conlon, and Gordy Kane for offering insightful suggestions on earlier drafts of this note, and to Herman Verlinde, Cumrun Vafa, Malcolm Perry, Jared Kaplan, and Piyush Kumar for stimulating discussions. This work was supported by the hospitality of the Cook's Branch Conservatory, and funded in part by a Graduate Research Fellowship from the National Science Foundation.

\begin{table*}[t=h]\caption{The complete branching of the adjoint of $E_8$ into its subgroup \mbox{$SU_3\times SU_2\times U_1^a\times U_1^b\times U_1^c\times U_1^d\times U_1^Y$}, where we have parameterized the $U_1$-factors according to standard conventions for $E_6$-like grand unified models---for example, $U_1^c$ is the Peccei-Quinn symmetry. Notice that we have only included the `un-barred' representations $\mathbf{R}_{\vec{q}\,\,}$---those for which $\vec{q}\in Q^+$; we apologize that this convention sets $\bar{\mathbf{27}}_{1,1}$ to be the `un-barred' representation $\mathbf{R}_{1,1}$ (while $\mathbf{27}_{\text{-}1,\text{-}1}\equiv\bar{\mathbf{R}}_{\text{-}1,\text{-}1}$ because it has `negative' $U_1$-charges). In the language of Ref.\ \cite{Bourjaily:2009vf}, this spectrum of matter singularities and corresponding interactions would be present in a manifold built out of the fibration \mbox{$\widehat{E_8}(a+c-d+Y,b+c-d+Y,-3c-d+Y,4d+Y,-5Y,0,0,0)$.\label{E8_branching}}}
\begin{tabular}{|llll@{\hspace{-0.5em}}r|crrrrrc|}\hline
$E_8\!\!\longrightarrow$&$\!\!\!\!\!\!\!\!\!\!E_6\!\times\! U_1^a\!\times\! U_1^b\!\!\to$&$\!\!\!\!\!SO_{10}\!\times\! U_1^c\!\!\to$&\multicolumn{2}{l}{$\!\!SU_5\times U_1^d\!\!\longrightarrow\!\!$}&$SU_3\!\times\! SU_2\!\!\!$&$\times U_1^a\!\!\!$&$\times U_1^b\!\!\!$&$\times U_1^c\!\!\!$&$\times U_1^d\!\!\!$&$\times U_1^Y\!\!\!\!\!$&\hspace{0.25em}{~}\\
\hline
\multirow{48}{*}{$\mathbf{248}\raisebox{34.05em}{~}\hspace{0.5em}\left\{\raisebox{25.25em}{$~$}\right.$}&\multirow{11}{*}{$\mathbf{27}_{1,0}\raisebox{9.75em}{~}\hspace{1em}\left\{\raisebox{6.05em}{$~$}\right.$}
&\multirow{6}{*}{$\mathbf{16}_1\raisebox{4.75em}{~}\hspace{1em}\left\{\raisebox{3.7em}{$~$}\right.$}&\multirow{3}{*}{$\mathbf{10}_{\text{-}1}\hspace{1em}\left\{\raisebox{2em}{$~$}\right.$}&$Q_1$&$(\mathbf{3},\bar{\mathbf{2}})$&1&0&1&-1&1&\\
&&&&$u^c_1$&$(\bar{\mathbf{3}},\mathbf{1})$&1&0&1&-1&-4&\\
&&&&$e^c_1$&$(\mathbf{1},\mathbf{1})$&1&0&1&-1&6&\\
&&&\multirow{2}{*}{$\bar{\mathbf{5}}_{3}\hspace{1.95em}\left\{\raisebox{1.5em}{$~$}\right.$}&$d^c_1$&($\bar{\mathbf{3}},\mathbf{1})$&1&0&1&3&2&\\
&&&&$L_1$&$(\mathbf{1},\mathbf{2})$&1&0&1&3&-3&\\
&&&$\mathbf{1}_{\text{-}5}$&$\nu^c_1$&$(\mathbf{1},\mathbf{1})$&1&0&1&-5&0&\\
&&\multirow{4}{*}{$\mathbf{10}_{\text{-}2}\raisebox{2.95em}{~}\hspace{0.725em}\left\{\raisebox{2.2em}{$~$}\right.$}&\multirow{2}{*}{$\mathbf{5}_2\hspace{1.95em}\left\{\raisebox{1.5em}{$~$}\right.$}&$D_1$&$(\mathbf{3},\mathbf{1})$&1&0&-2&2&-2&\\
&&&&$H^u_1$&$(\mathbf{1},\bar{\mathbf{2}})$&1&0&-2&2&3&\\
&&&\multirow{2}{*}{$\bar{\mathbf{5}}_{\text{-}2}\hspace{1.7em}\left\{\raisebox{1.5em}{$~$}\right.$}&$D_1^c$&$(\bar{\mathbf{3}},\mathbf{1})$&1&0&-2&-2&2&\\
&&&&$H^d_1$&$(\mathbf{1},\mathbf{2})$&1&0&-2&-2&-3&\\
&&$\mathbf{1}_4$&$\mathbf{1}_0$&$S_1$&$(\mathbf{1},\mathbf{1})$&1&0&4&0&0&\\

&\multirow{11}{*}{$\mathbf{27}_{0,1}\raisebox{9.75em}{~}\hspace{1em}\left\{\raisebox{6.05em}{$~$}\right.$}&\multirow{6}{*}{$\mathbf{16}_1\raisebox{4.75em}{~}\hspace{1em}\left\{\raisebox{3.7em}{$~$}\right.$}&\multirow{3}{*}{$\mathbf{10}_{\text{-}1}\hspace{1em}\left\{\raisebox{2em}{$~$}\right.$}&$Q_2$&$(\mathbf{3},\bar{\mathbf{2}})$&0&1&1&-1&1&\\
&&&&$u^c_2$&$(\bar{\mathbf{3}},\mathbf{1})$&0&1&1&-1&-4&\\
&&&&$e^c_2$&$(\mathbf{1},\mathbf{1})$&0&1&1&-1&6&\\
&&&\multirow{2}{*}{$\bar{\mathbf{5}}_{3}\hspace{1.95em}\left\{\raisebox{1.5em}{$~$}\right.$}&$d^c_2$&($\bar{\mathbf{3}},\mathbf{1})$&0&1&1&3&2&\\
&&&&$L_2$&$(\mathbf{1},\mathbf{2})$&0&1&1&3&-3&\\
&&&$\mathbf{1}_{\text{-}5}$&$\nu^c_2$&$(\mathbf{1},\mathbf{1})$&0&1&1&-5&0&\\
&&\multirow{4}{*}{$\mathbf{10}_{\text{-}2}\raisebox{2.95em}{~}\hspace{0.725em}\left\{\raisebox{2.2em}{$~$}\right.$}&\multirow{2}{*}{$\mathbf{5}_2\hspace{1.95em}\left\{\raisebox{1.5em}{$~$}\right.$}&$D_2$&$(\mathbf{3},\mathbf{1})$&0&1&-2&2&-2&\\
&&&&$H^u_2$&$(\mathbf{1},\bar{\mathbf{2}})$&0&1&-2&2&3&\\
&&&\multirow{2}{*}{$\bar{\mathbf{5}}_{\text{-}2}\hspace{1.7em}\left\{\raisebox{1.5em}{$~$}\right.$}&$D_2^c$&$(\bar{\mathbf{3}},\mathbf{1})$&0&1&-2&-2&2&\\
&&&&$H^d_2$&$(\mathbf{1},\mathbf{2})$&0&1&-2&-2&-3&\\
&&$\mathbf{1}_4$&$\mathbf{1}_0$&$S_2$&$(\mathbf{1},\mathbf{1})$&0&1&4&0&0&\\

&\multirow{11}{*}{$\bar{\mathbf{27}}_{1,1}\raisebox{9.75em}{~}\hspace{1em}\left\{\raisebox{6.05em}{$~$}\right.$}&\multirow{6}{*}{$\bar{\mathbf{16}}_{\text{-}1}\raisebox{4.75em}{~}\hspace{0.75em}\left\{\raisebox{3.7em}{$~$}\right.$}&\multirow{3}{*}{$\bar{\mathbf{10}}_{1}\hspace{1.25em}\left\{\raisebox{2em}{$~$}\right.$}&$\bar{Q_3}$&$(\bar{\mathbf{3}},\mathbf{2})$&1&1&-1&1&-1&\\
&&&&$\bar{u_3^c}$&$(\mathbf{3},\mathbf{1})$&1&1&-1&1&4&\\
&&&&$\bar{e_3^c}$&$(\mathbf{1},\mathbf{1})$&1&1&-1&1&-6&\\
&&&\multirow{2}{*}{$\mathbf{5}_{\text{-}3}\hspace{1.7em}\left\{\raisebox{1.5em}{$~$}\right.$}&$\bar{d_3^c}$&($\mathbf{3},\mathbf{1})$&1&1&-1&-3&-2&\\
&&&&$\bar{L_3}$&$(\mathbf{1},\bar{\mathbf{2}})$&1&1&-1&-3&3&\\
&&&$\mathbf{1}_{5}$&$\bar{\nu_3^c}$&$(\mathbf{1},\mathbf{1})$&1&1&-1&5&0&\\
&&\multirow{4}{*}{$\mathbf{10}_{\phantom{\text{-}}2}\raisebox{2.95em}{~}\hspace{0.725em}\left\{\raisebox{2.2em}{$~$}\right.$}&\multirow{2}{*}{$\bar{\mathbf{5}}_{\text{-}2}\hspace{1.7em}\left\{\raisebox{1.5em}{$~$}\right.$}&$\bar{D_3}$&$(\bar{\mathbf{3}},\mathbf{1})$&1&1&2&-2&2&\\
&&&&$\bar{H^u_3}$&$(\mathbf{1},\mathbf{2})$&1&1&2&-2&-3&\\
&&&\multirow{2}{*}{$\mathbf{5}_{2}\hspace{1.95em}\left\{\raisebox{1.5em}{$~$}\right.$}&$\bar{D_3^c}$&$(\mathbf{3},\mathbf{1})$&1&1&2&2&-2&\\
&&&&$\bar{H^d_3}$&$(\mathbf{1},\bar{\mathbf{2}})$&1&1&2&2&3&\\
&&$\mathbf{1}_{\text{-}4}$&$\mathbf{1}_0$&$\bar{S_3}$&$(\mathbf{1},\mathbf{1})$&1&1&-4&0&0&\\

&$\mathbf{1}_{2,1}$&$\mathbf{1}_{0}$&$\mathbf{1}_{0}$&$N_1$&$(\mathbf{1},\mathbf{1})$&2&1&0&0&0&\\
&$\mathbf{1}_{1,2}$&$\mathbf{1}_{0}$&$\mathbf{1}_{0}$&$N_2$&$(\mathbf{1},\mathbf{1})$&1&2&0&0&0&\\
&$\mathbf{1}_{1,\text{-}1}$&$\mathbf{1}_{0}$&$\mathbf{1}_{0}$&$N_3$&$(\mathbf{1},\mathbf{1})$&1&-1&0&0&0&\\
&\multirow{10}{*}{$\!(\mathbf{78}_{0,0})\raisebox{8.25em}{~}\hspace{0.4em}\left\{\raisebox{4.45em}{$~$}\right.$}&\multirow{6}{*}{$\bar{\mathbf{16}}_{3}\raisebox{4.75em}{~}\hspace{1em}\left\{\raisebox{3.7em}{$~$}\right.$}&\multirow{3}{*}{$\bar{\mathbf{10}}_{1}\hspace{1.25em}\left\{\raisebox{2em}{$~$}\right.$}&$\bar{Q_4}$&$(\bar{\mathbf{3}},\mathbf{2})$&0&0&3&1&-1&\\
&&&&$\bar{u_4^c}$&$(\mathbf{3},\mathbf{1})$&0&0&3&1&4&\\
&&&&$\bar{e_4^c}$&$(\mathbf{1},\mathbf{1})$&0&0&3&1&-6&\\
&&&\multirow{2}{*}{$\mathbf{5}_{\text{-}3}\hspace{1.7em}\left\{\raisebox{1.5em}{$~$}\right.$}&$\bar{d_4^c}$&$(\mathbf{3},\mathbf{1})$&0&0&3&-3&-2&\\
&&&&$\bar{L_4}$&$(\mathbf{1},\bar{\mathbf{2}})$&0&0&3&-3&3&\\
&&&$\mathbf{1}_{5}$&$\bar{\nu_4^c}$&$(\mathbf{1},\mathbf{1})$&0&0&3&5&0&\\
&&\multirow{4}{*}{$\!(\mathbf{45}_{0})\raisebox{3.5em}{~}\hspace{0.365em}\left\{\raisebox{2em}{$~$}\right.$}&\multirow{3}{*}{$\mathbf{10}_{4}\hspace{1.25em}\left\{\raisebox{2em}{$~$}\right.$}&$Q_5$&$(\mathbf{3},\bar{\mathbf{2}})$&0&0&0&4&1&\\
&&&&$u_5^c$&$(\bar{\mathbf{3}},\mathbf{1})$&0&0&0&4&-4&\\
&&&&$e_5^c$&$(\mathbf{1},\mathbf{1})$&0&0&0&4&6&\\
&&&\multirow{1}{*}{$\!(\mathbf{24}_{0})$}&$Q_Y^c$&$(\bar{\mathbf{3}},\mathbf{2})$&0&0&0&0&5&\\\hline
\end{tabular}
\end{table*}

\newpage%~\newpage~\newpage
%\bibliographystyle{nhieeetr}
%\bibliography{refs}

\begin{thebibliography}{10}

\bibitem{Beasley:2008dc}
C.~Beasley, J.~J. Heckman, and C.~Vafa, ``{GUTs and Exceptional Branes in
  F-theory - I},'' 2008, arXiv:0802.3391~[hep-th].

\bibitem{Beasley:2008kw}
C.~Beasley, J.~J. Heckman, and C.~Vafa, ``{GUTs and Exceptional Branes in
  F-theory - II: Experimental Predictions},'' 2008, arXiv:0806.0102~[hep-th].

\bibitem{Bourjaily:2009vf}
J.~L. Bourjaily, ``{Local Models in F-Theory and M-Theory with Three
  Generations},'' 2009, arXiv:0901.3785~[hep-th].

\bibitem{Heckman:2008jy}
J.~J. Heckman, A.~Tavanfar, and C.~Vafa, ``{Cosmology of F-theory GUTs},''
  2008, arXiv:0812.3155~[hep-th].

\bibitem{Heckman:2008ads}
J.~J. Heckman and C.~Vafa, ``{F-Theory, GUTs, and the Weak Scale},'' 2008,
  arXiv:0809.1098~[hep-th].

\bibitem{Marsano:2008jq}
J.~Marsano, N.~Saulina, and S.~Schafer-Nameki, ``{Gauge Mediation in F-Theory
  GUT Models},'' 2008, arXiv:0808.1571~[hep-th].

\bibitem{Heckman:2008qa}
J.~J. Heckman and C.~Vafa, ``{Flavor Hierarchy from F-theory},'' 2008,
  arXiv:0811.2417~[hep-th].

\bibitem{Blumenhagen:2008aw}
R.~Blumenhagen, ``{Gauge Coupling Unification in F-Theory Grand Unified
  Theories},'' {\em Phys. Rev. Lett.}, vol.~102, p.~071601, 2009,
  arXiv:0812.0248~[hep-th].

\bibitem{Donagi:2008ca}
R.~Donagi and M.~Wijnholt, ``{Model Building with F-Theory},'' 2008,
  arXiv:0802.2969~[hep-th].

\bibitem{Donagi:2008kj}
R.~Donagi and M.~Wijnholt, ``{Breaking GUT Groups in F-Theory},'' 2008,
  arXiv:0808.2223~[hep-th].

\bibitem{Wijnholt:2008db}
M.~Wijnholt, ``{F-Theory, GUTs and Chiral Matter},'' 2008,
  arXiv:0809.3878~[hep-th].

\bibitem{Heckman:2009bi}
J.~J. Heckman, G.~L. Kane, J.~Shao, and C.~Vafa, ``{The Footprint of F-theory
  at the LHC},'' 2009, arXiv:0903.3609~[hep-ph].

\bibitem{Donagi:2009ra}
R.~Donagi and M.~Wijnholt, ``{Higgs Bundles and UV Completion in F-Theory},''
  2009, arXiv:0904.1218.

\bibitem{Marsano:2009ym}
J.~Marsano, N.~Saulina, and S.~Schafer-Nameki, ``{F-theory Compactifications
  for Supersymmetric GUTs},'' 2009, arXiv:0904.3932~[hep-th].

\bibitem{Bouchard:2009bu}
V.~Bouchard, J.~J. Heckman, J.~Seo, and C.~Vafa, ``{F-Theory and Neutrinos:
  Kaluza-Klein Dilution of Flavor Hierarchy},'' 2009, arXiv:0904.1419~[hep-ph].

\bibitem{Acharya:2001gy}
B.~S. Acharya and E.~Witten, ``Chiral {F}ermions from {M}anifolds of ${G}_2$
  {H}olonomy,'' 2001, hep-th/0109152.

\bibitem{Acharya:2004qe}
B.~S. Acharya and S.~Gukov, ``{M Theory and Singularities of Exceptional
  Holonomy Manifolds},'' {\em Phys. Rept.}, vol.~392, pp.~121--189, 2004,
  hep-th/0409191.

\bibitem{Berglund:2002hw}
P.~Berglund and A.~Brandhuber, ``Matter from ${G}_2$ {M}anifolds,'' {\em Nucl.
  Phys.}, vol.~B641, pp.~351--375, 2002, hep-th/0205184.

\bibitem{Bourjaily:2008ji}
J.~L. Bourjaily and S.~Espahbodi, ``{Geometrically Engineerable Chiral Matter
  in M-Theory},'' 2008, arXiv:0804.1132~[hep-th].

\bibitem{Bourjaily:2007kv}
J.~L. Bourjaily, ``{Unfolding Geometric Unification in M-Theory},'' 2007,
  arXiv:0706.3364~[hep-th].

\bibitem{Katz:1996xe}
S.~Katz and C.~Vafa, ``Matter from {G}eometry,'' {\em Nucl. Phys.}, vol.~B497,
  pp.~146--154, 1997, hep-th/9606086.

\bibitem{Witten:2001uq}
E.~Witten, ``Anomaly {C}ancellation on ${G}_2$ {M}anifolds,'' 2001,
  hep-th/0108165.

\bibitem{Cvetic:2001nr}
M.~Cvetic, G.~Shiu, and A.~M. Uranga, ``{Chiral Four-Dimensional
  $\mathcal{N}=1$ Supersymmetric Type IIA Orientifolds from Intersecting
  D6-Branes},'' {\em Nucl. Phys.}, vol.~B615, pp.~3--32, 2001, hep-th/0107166.

\bibitem{FultonRepTheory}
W.~Fulton and J.~Harris, {\em Representation Theory}.
\newblock Graduate Texts in Mathematics, Springer, 2000.

\bibitem{Slansky:1981yr}
R.~Slansky, ``Group {T}heory for {U}nified {M}odel {B}uilding,'' {\em Phys.
  Rept.}, vol.~79, pp.~1--128, 1981.

\bibitem{Atiyah:2001qf}
M.~Atiyah and E.~Witten, ``M-{T}heory {D}ynamics on a {M}anifold of ${G}_2$
  {H}olonomy,'' {\em Adv. Theor. Math. Phys.}, vol.~6, pp.~1--106, 2003,
  hep-th/0107177.

\bibitem{Bourjaily:2009aa}
J.~L. Bourjaily and M.~J. Perry 2009.
\newblock Work in progress.

\bibitem{Aspinwall:1995vk}
P.~S. Aspinwall and J.~Louis, ``{On the Ubiquity of $K3$ Fibrations in String
  Duality},'' {\em Phys. Lett.}, vol.~B369, pp.~233--242, 1996, hep-th/9510234.

\bibitem{Bourjaily:2007vx}
J.~L. Bourjaily, ``{Geometrically Engineering the Standard Model: Locally
  Unfolding Three Families out of ${E}_8$},'' {\em Phys. Rev.}, vol.~D76,
  p.~046004, 2007, arXiv:0704.0445~[hep-th].

\bibitem{Langacker:1980js}
P.~Langacker, ``{Grand Unified Theories and Proton Decay},'' {\em Phys. Rept.},
  vol.~72, p.~185, 1981.

\bibitem{Hewett:1988xc}
J.~L. Hewett and T.~G. Rizzo, ``{Low-Energy Phenomenology of Superstring
  Inspired $E(6)$ Models},'' {\em Phys. Rept.}, vol.~183, p.~193, 1989.

\bibitem{Ibe:2007km}
M.~Ibe and R.~Kitano, ``{Sweet Spot Supersymmetry},'' {\em JHEP}, vol.~08,
  p.~016, 2007, arXiv:0705.3686~[hep-ph].

\bibitem{Giudice:1988yz}
G.~F. Giudice and A.~Masiero, ``{A Natural Solution to the $\mu$ Problem in
  Supergravity Theories},'' {\em Phys. Lett.}, vol.~B206, pp.~480--484, 1988.

\bibitem{Froggatt:1978nt}
C.~D. Froggatt and H.~B. Nielsen, ``{Hierarchy of Quark Masses, Cabibbo Angles
  and CP Violation},'' {\em Nucl. Phys.}, vol.~B147, p.~277, 1979.

\bibitem{Bilal:2003bf}
A.~Bilal and S.~Metzger, ``{Compact Weak $G(2)$-Manifolds with Conical
  Singularities},'' {\em Nucl. Phys.}, vol.~B663, pp.~343--364, 2003,
  hep-th/0302021.

\bibitem{Acharya:2006ia}
B.~S. Acharya, K.~Bobkov, G.~Kane, P.~Kumar, and D.~Vaman, ``{An M-Theory
  Solution to the Hierarchy Problem},'' {\em Phys. Rev. Lett.}, vol.~97,
  p.~191601, 2006, hep-th/0606262.

\bibitem{Acharya:2008zi}
B.~S. Acharya, K.~Bobkov, G.~L. Kane, J.~Shao, and P.~Kumar, ``{The
  $G_2$-MSSM-An M-Theory Motivated Model of Particle Physics},'' 2008,
  arXiv:0801.0478 \mbox{[hep-ph]}.

\bibitem{Acharya:2007rc}
B.~S. Acharya, K.~Bobkov, G.~L. Kane, P.~Kumar, and J.~Shao, ``{Explaining the
  Electroweak Scale and Stabilizing Moduli in M-Theory},'' {\em Phys. Rev.},
  vol.~D76, p.~126010, 2007, hep-th/0701034.

\bibitem{Chung:2003fi}
D.~J.~H. Chung, L.~L. Everett, G.~L. Kane, S.~F. King, J.~D. Lykken, and L.-T.
  Wang, ``{The Soft Supersymmetry-Breaking Lagrangian: Theory and
  Applications},'' {\em Phys. Rept.}, vol.~407, pp.~1--203, 2005,
  hep-ph/0312378.

\bibitem{Conlon:2008cj}
J.~P. Conlon, R.~Kallosh, A.~Linde, and F.~Quevedo, ``{Volume Modulus Inflation
  and the Gravitino Mass Problem},'' {\em JCAP}, vol.~0809, p.~011, 2008,
  arXiv:0806.0809~[hep-th].

\bibitem{Conlon:2008wa}
J.~P. Conlon, A.~Maharana, and F.~Quevedo, ``{Towards Realistic String
  Vacua},'' 2008, arXiv:0810.5660~[hep-th].

\bibitem{Preskill:1990fr}
J.~Preskill, ``{Gauge Anomalies in an Effective Field Theory},'' {\em Ann.
  Phys.}, vol.~210, pp.~323--379, 1991.

\end{thebibliography}

\end{document}